\documentclass[structabstract]{aa}
\usepackage{graphicx}
\usepackage{txfonts}
\usepackage{color}
\usepackage{natbib}

\newcommand{\farcss}{\mbox{\rlap{.}$''$}}

\newcommand{\kms}{km~s$^{-1}$}

\begin{document}
%

   \title{Modeling the physical and excitation conditions of the molecular envelope of NGC~7027\thanks{Based on
observations made with the Herschel Space Telescope operated by ESA and the 30-m radiotelescope operated by IRAM.}}


\author{M. Santander-Garc\'\i a\inst{1,2} 
    \and V. Bujarrabal\inst{1}
    \and J. Alcolea\inst{1}
            }

\institute{
        Observatorio Astron\'omico Nacional, Ap.\ de Correos 112, E-28803, Alcal\'a de Henares, Madrid, Spain \\ email: {\tt m.santander@oan.es}
	\and
	Centro de Astrobiolog\'\i a, CSIC-INTA, Ctra de Torrej\'on a Ajalvir km 4, E-28850 Torrej\'on de Ardoz, Spain
        }



  \abstract
   {The link between the shaping of bipolar planetary nebulae and the mass ejection activity of their central stars is still poorly understood. Appropriately characterizing the evolution of the shells ejected during the late stages of stellar evolution and the interaction between these shells is  fundamental to gain insight into the mechanism of nebular shaping. This must include the study of the molecular emission, which tracks the mass-loss history during the late AGB and post-AGB stages, when the nebula is being actively shaped.}
   {Herschel/HIFI provides an invaluable tool by opening a new window (most of the sub-mm and far infrared range is only accessible from space) from which to probe warm molecular gas ($\sim$50--1000 K). This paper presents a radiative-transfer, spatio-kinematic modeling of the molecular envelope of the young planetary nebula NGC~7027 in several high- and low$-J$ $^{12}$CO and $^{13}$CO transitions observed by Herschel/HIFI and the IRAM 30-m radio-telescope, and discusses the structure and dynamics of the molecular envelope.}
   {We have developed a code which, used along with the existing SHAPE software (\citealp{steffen11}), implements spatio-kinematic modeling with accurate non-LTE calculations of line excitation and radiative transfer in molecular species. We have used this code to build a relatively simple ``russian doll'' model to account for the physical and excitation conditions of the molecular envelope of NGC~7027.}
   {The model nebula consists of four nested, mildly bipolar shells plus a pair of high-velocity blobs. The innermost shell is the thinnest and shows a significant jump in physical conditions (temperature, density, abundance and velocity) with respect to the adjacent shell. This is a clear indication of a shock front in the system, which may have played a role in the shaping of the nebula. Each of the high-velocity blobs is divided into two sections with considerably different physical conditions. The striking  presence of H$_2$O in NGC~7027, a C-rich nebula, is likely due to photo-induced chemistry from the hot central star, although formation of water by shocks cannot be ruled out. The computed molecular mass of the nebula is 1.3 M$_\odot$, compatible with that derived from previous works.}
   {}


   \maketitle
%

\section{Introduction}

The physical mechanisms behind the shaping of proto-planetary nebulae (PPNe) and young planetary nebulae (PNe) has been and still is a subject of intense debate (see e.g. \citealp{balick02}). In particular, PPNe and young PNe usually show high-velocity outflows, usually in the polar directions, along with slower components whose velocities are comparable to those of the winds of AGB envelopes. These fast-bipolar outflows are thought to be the result of the propagation of shocks in the polar regions of the dense AGB shell during the evolution towards PPNe. However, many aspects of this process are poorly understood at present.

Tracking the mass ejecta of planetary nebulae (PNe) and proto-planetary nebulae (PPNe) by means of molecular emission lines can provide crucial clues for better understanding this topic. Herschel/HIFI provides an invaluable tool, as it probes warm molecular gas ($\sim$50--1000~K) by producing high-resolution spectra in high-excitation transitions (e.g. $^{12}$CO $J$=6$-$5, 10$-$9 and 16$-$15), some of which are unobservable from the ground, rendering ground-based telescopes esentially useless to accurately probe molecular gas with temperatures over $\sim$100~K.

Recently, as part of the guaranteed-time key program HIFISTARS, a sample of ten PPNe and young PNe were observed with Herschel/HIFI (\citealp{bujarrabal11}). Although HIFISTARS only includes observations towards the center of these nebulae, the kinematics and the excitation conditions of the warm molecular gas of the target can be studied with unprecedented detail thanks to the high spectral resolution of the profiles.

This work focuses on one of the targets observed within the HIFISTARS program, NGC~7027 (PN G084.9-03.4). With a kinematical age of just 600 years (based on its radio flux, see \citealt{masson89}) and located at 1~kpc (\citealp{zijlstra08}), this compact and young PNe is one of the brightest nebulae in the sky and the most extensively studied PNe so far. NGC~7027 is a carbon-rich nebula with a very high-excitation spectrum (e.g. it shows ion lines such as O {\sc iv} and Mg {\sc v}). It hosts one of the hottest central stars known to date, with a T$_\mathrm{eff}\sim$200,000 K (e.g \citealp{latter00}). A small, essentially ellipsoidal, expanding ionized shell surrounds the central star (\citealp{masson89}). Further outwards, a thin H$_2$ shell indicates the presence of a photo-dissociation region (PDR) and shows signs of recent interaction with collimated outflows (\citealp{cox02}). The nebula also shows molecular emission from CO beyond the PDR, extending further out to 15-20 arcsec from the central star. In particular, \cite{nakashima10}  carried out interferometric observations in the CO $J$=2$-$1 and 1$-$0 lines and found that the CO envelope could be roughly modeled as an ellipsoidal shell, despite some deviations caused by a high-velocity component. The kinematics of this cold, molecular envelope seems to be closely linked with that of the inner ionized shell and the PDR (\citealp{cox97}, \citealp{latter00}).

The amount of molecular mass inferred from the CO emission is very significant. In fact, despite hosting one of the hottest central stars known, the ionized mass only represents 2\% of the total measured mass of the nebula, 1.4 M$_\odot$, while the vast majority of the nebula (85\%) consists of molecular gas, the remaining 13\% being in the form of neutral atomic gas (\citealp{fong01}). It seems clear that any approach towards understanding the shaping mechanism of this nebula must take into account the emission from molecular gas.

In this work, we present a detailed study of the molecular envelope of NGC~7027, combining observational data from low-$J$ transitions, observed with the IRAM 30-m radio-telescope with data from high-$J$, observed as part of the HIFISTARS program. We have developed a code, {\tt shapemol}, which, used along with {\tt SHAPE} (\citealp{steffen11}), implements spatio-kinematic modeling with accurate calculations of non-LTE line excitations and radiative transfer in molecular species. The high quality of the HIFI data, together with {\tt shapemol}, have allowed us to study, for the first time in such level of detail, the kinematics and excitation conditions of the warm, molecular (CO-rich) gas of a PN.

A preliminar version of this work was reported in \cite{santander11}.

\section{Observations}

The observations used in this paper (see Fig.~\ref{F1}) for the modeling of the molecular
envelope of NGC~7027 have been already presented by \cite{bujarrabal01}, $^{12}$CO and
$^{13}$CO $J$=1$-$0 and 
2$-$1 data, and \cite{bujarrabal11}, rest of the data consisting of the high-$J$ transitions.  Here we will
briefly summarize the 
main observational characteristics of the data (see Table 1), as they have been
described in full detail 
in these papers. CO $J$=1$-$0 and 2$-$1 spectra were obtained in 1999
and 2000 using the IRAM 
30-m at Pico de Veleta (Spain). This is a very well characterized
ground-based instrument, 
in which the main source of uncertainty in the absolute calibration of
the data comes from the 
correction for the opacity of the atmosphere. Although this effect is
measured every 20 to 30 
min, in standard conditions we cannot guarantee accuracies better than
20--25\% when comparing 
data from Pico de Veleta with those from other telescopes, as this is the best accuracy our group has found on long-term observations of the same standard sources, specially for observations dating a decade or more. Spectra at
higher frequencies are taken 
from the Herschel/HIFI HIFISTARS GTKP (P.I. V.~Bujarrabal). In this case,
the main source or 
uncertainty in the absolute calibration of the data of this space-based
telescope comes from the 
fact that the gain ratio between the two sidebands of the receivers is not
well known. For the 
HIFISTARS data we have adopted uncertainty values from 15\% to 30\%
depending on the receiver
band, following the description of the HIFI instrument characteristics
and performances given by
\cite{roelfsema12}. In all cases we have used the $T_\mathrm{mb}$ intensity
scale, that can be
directly compared with the model results when these take into account
the shape of the 
beam of the telescope (see below).

In all cases, the spectra are taken towards the central star, located at J2000 R.A. 21$^ \mathrm{h}$07$^ \mathrm{m}$01$^ \mathrm{s}$.59 Dec.$+$42$^\circ$14$'$10\farcss 2. 
The nebula is
sometimes larger than
the beam of the telescope (12--22 arcsec for the 30-m, and 12--33
arcsec for HIFI), 
and therefore by observing just the central position we do not
observe its full emission. This is particularly true for the higher
frequencies in both
telescopes (at 220--230 GHz for IRAM 30-m and 1800 GHz for
Herschel/HIFI), but this effect
has been properly taken into account in our modeling (see section 4).
For this purpose
we have assumed that the telescope beam shape is well represented by a
symmetrical 2D-Gaussian
solely characterized by its FWHP, what has been taken to be equal to the
measured HPBW of
the two instruments at the corresponding frequencies (see Table 1). 

The spectral resolution of the instruments is also limited. For the IRAM
30-m observations, 
the data were taken from the (now decommissioned) 100~kHz filter-bank,
after degrading the 
resolution to 300~kHz by averaging every three channels, resulting in
velocity resolutions of 
0.8--0.9~km~s$^{-1}$ for the 1--0 lines, and of 0.4~km~s$^{-1}$ for the 2--1 lines. For the
Herschel/HIFI data, we have 
used the Wide Band Spectrometer (WBS) in standard configuration,
providing a spectral 
resolution of about 1.1~MHz, slightly varying across the observed band.
The data have been 
oversampled in a homogeneous  0.5~MHz spectral channels, that we have
degraded down to 
resolutions equivalent to 0.4 to 1.6~km~s$^{-1}$) depending of the
strength of the detected 
line and the required S/N ratio.

\begin{table*}
  \begin{center}
  \caption{Observing log and profile features.}
  \label{T1}
 {\scriptsize
  \begin{tabular}{|l|c|c|c|c|c|c|c|}
\hline 
{\bf Species} & {\bf Transition} &  {\bf Frequency} & {\bf Resolution} & {\bf T$_{\mathrm{mb}}$ peak$^\dagger$} & {\bf r.m.s.} & {\bf Integ. flux$^\star$} & {\bf HPBW} \\
  &   & {\bf (GHz)} & {\bf (km~s$^{-1}$)} & {\bf (K)} & {\bf (K)} & {\bf (K~km~s$^{-1}$)} & {\bf ($''$)} \\
\hline
\hline
\multicolumn{7}{|l|}{IRAM 30-m observations$^{\mathrm a}$} \\
\hline
$^{12}$CO & $J$=1$-$0 & 115.271 &  0.8    & 11.9  &  7.7$\times$10$^{-2}$  & 332  & 22 \\
$^{12}$CO & $J$=2$-$1 & 230.538 &  0.4   &  30.9 &  9.9$\times$10$^{-2}$  & 667  & 12 \\
$^{13}$CO & $J$=1$-$0 & 110.201 &  0.9    & 0.3  &  9.4$\times$10$^{-3}$  & 7.9  & 22 \\
$^{13}$CO & $J$=2$-$1 & 220.399 &  0.4    & 1.3  &  3.0$\times$10$^{-2}$  & 31  & 12 \\
\hline\hline
\multicolumn{7}{|l|}{Herschel/HIFI observations$^{\mathrm b}$} \\
\hline
$^{12}$CO & $J$=6$-$5 & 691.473 &  0.4    & 5.1  &  1.3$\times$10$^{-2}$  & 137  & 31 \\
$^{12}$CO & $J$=10$-$9 & 1151.985 &  0.4    & 5.5  &  5.5$\times$10$^{-2}$  & 146  & 20 \\
$^{12}$CO & $J$=16$-$15 & 1841.346 &  0.5    & 2.5  &  9.3$\times$10$^{-2}$  & 52  & 12 \\
$^{13}$CO & $J$=6$-$5 & 661.067  &  0.5    & 0.3  &  8.6$\times$10$^{-3}$  & 8.3  & 31 \\
$^{13}$CO & $J$=10$-$9 & 1101.350 &  0.5    & 0.2  &  2.5$\times$10$^{-2}$  & 5.7  & 20 \\
C$^{18}$O & $J$=6$-$5 & 658.506 &  0.9    & 0.03  &  6.1$\times$10$^{-3}$  & 0.7  & 33 \\
OH & $^2\Pi_{1/2}$ $J$=3/2$^-$-1/2$^+$ & 1834.75 &  1.6    & 0.16  &  4.8$\times$10$^{-2}$  & 6.3  & 12 \\
p-H$_ 2$O & $J_{K_a,K_c}$=$1_{1,1}-0_{0,0}$  & 1113.343 & 0.9    & 0.1  &  1.9$\times$10$^{-2}$  & 2.1  & 20 \\
o-H$_ 2$O & $J_{K_a,K_c}$=$1_{1,0}-1_{0,1}$ & 556.936 &  0.5    & 0.1  &  5.5$\times$10$^{-3}$  & 2.1  & 39 \\
\hline
  \end{tabular}
  }
 \end{center}
\vspace{1mm}
  \scriptsize{
  {\it Notes:}\\
   $^\dagger$Maximum intensity of the profile in main beam temperature $T_{\mathrm{mb}}$(K), defined as the peak intensity minus the r.m.s. in the spectrum \\
   $^\star$Integrated flux of the profile.  \\
   $^{\mathrm a}$ Observations published in \cite{bujarrabal01} \\
   $^{\mathrm b}$ Observations published in \cite{bujarrabal11}
}
\end{table*}

\section{Modeling}

The main goal of our modeling is to determine the physical conditions (densities, temperatures and abundances) and the general kinematic structure of the nebula. Note that an accurate determination of the 3-D structure from our data is prevented by the lack of spatial resolution. In fact, the only information about the spatial structure is concealed in the frequency-dependent beam size, which weighs the emission from the central region and that coming from the outer region in a different fashion for each transition, and which is, for the highest-$J$ transitions, smaller than the molecular CO shell at lower-$J$ described by previous works (see beam size column in Table \ref{T1}). Hence, the general geometry of the nebula has been deduced both from the analysis of previous works and a series of reasonable deductions from the Herschel and IRAM data-set (see section \ref{adopted}). In any case, the following model should be considered as a first, simplified approach to a much more complex reality. 

The modeling process is similar to standard spatio-kinematic modeling (e.g. \citealp{santander04b}), except that it incorporates full radiative-transfer solving for different transitions, creating synthetic profiles to be matched against the observations. We have achieved this by using a special version (4.51$\beta$) of the {\tt SHAPE}\footnote{{\tt SHAPE} is available on http://bufadora.astrosen.unam.mx/shape/} software by \cite{steffen11}, specifically customized for usage alongside our own {\tt GDL/IDL}-based code, {\tt shapemol}\footnote{If interested in using {\tt shapemol}, contact the first author.}. 

{\tt SHAPE} is a software tool for building spatio-kinematic models of nebulae. It allows to easily implement a 3-D structure and 3-D velocity field to describe the model nebula, and generate synthetic images, position-velocity diagrams and/or channel maps for direct comparison with observations. Its versatility has made it the standard tool in the field, where it has proven to be valuable for producing accurate spatio-kinematic descriptions of many planetary nebulae (e.g. \citealp{steffen11}, \citealp{jones10b}, \citealp{velazquez11}).

Although {\tt SHAPE} implements radiative-transfer solving for atomic species, in which the absorption and emission coefficients are easily predictable over a large range of temperatures, densities and abundances, it is unable, on its own, to work with molecular species, either in the thermalized or non-thermalized cases. The main reason behind this is the strong dependence of the absorption and emission coefficients of each grid cell, not only on the local abundance, temperature and density, but also on the size of the emitting structure and the velocity of the grid cell along the line of sight.

We designed {\tt shapemol} to fill this gap. In the following subsection, we describe this software and its role in the modeling process.

\subsection{{\tt shapemol}}
{\tt shapemol} is a GDL/IDL-based code built to be used as a
complement for {\tt SHAPE}, allowing for radiative transfer solving in
molecular transitions. In particular, it performs non-LTE calculations
of line excitations (based on the well-known LVG approximation, see below)
to compute the absorption ($\kappa_\nu$) and emission ($j_\nu$)
coefficients of each individual cell in the grid, for the desired
transition and species.

In spectral lines, the values of $\kappa_\nu$ and $j_\nu$ are given by
the populations of the involved energy levels and the line profile
shape. In our case (low-frequency transitions requiring relatively low
excitation), the profile shape is given by the local velocity
dispersion due to thermal dispersion or turbulence. The level
populations depend on the collisional transition rates and the
radiative excitation and de-excitation rates, which in turn depend
on the amount of radiation reaching the nebula point we are
considering at the frequency of the line (averaged over angle and
frequency within the local line profile). The calculation of such
averaged radiation intensity requires a previous knowledge of the
absorption and emission coefficients in the whole cloud, in order to
solve the radiative transfer equation in all directions and
frequencies. Of course, the populations of a high number of levels
must be calculated simultaneously, since the population of each one
depends on those of the others, and in all the points of the
cloud. So, the solution of the system becomes extremely complex in the
general case.

The problem is greatly simplified when there is a Large Velocity
Gradient (LVG) in the cloud, introducing important Doppler shifts
between points that are sufficiently far away. When this shift is
larger than the local velocity dispersions, the points cannot
radiatively interact at large scales, so the
radiative transfer is basically a local phenomenon. The excitation equations 
can be then solved, and $\kappa_\nu$ and
$j_\nu$ calculated in each point, independently of the rest of the cloud. 
In any case, the level populations depend in a complex way on the (local)
physical conditions, and the solution requires an iterative process. A
very understandable version of the general formalism can be seen in
\cite{castor70}. The LVG approximation includes the main ingredients of
the problem (collisional and radiative rates, trapping when opacities
are high, population transfer between different levels, etc.) and gives
fast, sensible solutions. These excitation calculations are quite
accurate, even when the LVG conditions are barely satisfied, at least
for the case of molecular lines from shells around evolved stars
(e.g. Bujarrabal et al. in preparation; \citealp{teyssier06};
\citealp{ramstedt08}). The approximation itself is not
necessary to derive the resulting line profiles,
which can be calculated solving the standard transfer equation using
the level populations derived from the LVG approximation and a local
velocity dispersion, as indeed we have done using {\tt SHAPE}.

According to the LVG approximation, $\kappa_\nu$ and $j_\nu$ depend in
a heavily non-linear way on the density $n$ and the temperature $T$,
and almost linearly on the product $\frac{r}{V} X$, where $r$ is the
distance of a given point (or code cell) to the central star, $V$ its
velocity and $X$ the abundance of the desired species. Besides these
three parameters, the LVG results depend on the logarithmic
velocity gradient, $\epsilon$ = d$V$/d$r$\,$r$/$V$, but only slightly. 
We will assume $\epsilon$ = 1, i.e.\ a linear dependence
of $V$ on $r$. In this case, the escape probability of a photon is given by a
simple analytical function of the opacity and its calculation is
considerably simplified. Such velocity fields are found in many young
planetary nebulae, which are basically expanding at high velocity
following a 'Hubble' velocity law.

The approach of {\tt shapemol} consists of relying on a set of
pre-generated tables of $\kappa_\nu$ and $j_\nu$ as functions of $n$
and $T$, each table corresponding to a value of the $\frac{r}{V} X$
product. The values of $\kappa_\nu$ and $j_\nu$ in each table are
individually computed for each pair of $n$ and $T$ by iteratively
solving a set of equations via the convergence algorithm typical of
LVG codes. In this calculations 20-25 rotational levels were taken
into account, allowing accurate calculations for temperatures up to at
least 1000 K, and we used collisional rates from the Leiden Atomic and
Molecular Database (LAMDA; see \citealp{schoier05},
http://www.strw.leidenuniv.nl/\~moldata/).  For a set of physical
conditions, {\tt shapemol} selects the table with the closest value of
$\frac{r}{V} X$. Once a table has been selected, {\tt shapemol}
computes $\kappa_\nu$ and $j_\nu$ by linear interpolation between the
values for the two adjacent tabulated values of $n$ and $T$. The steps
in $n$ and $T$ are small enough to guarantee that a 1$^\mathrm{st}$ order
interpolation is a good approximation between two consecutive
values. Finally, given the roughly linear dependence of $\kappa_\nu$
and $j_\nu$ on the product $\frac{r}{V} X$, the software scales the
computed absorption and emission coefficients according to the ratio
of the desired value of $\frac{r}{V} X$ to that of the selected table; to avoid
significant errors, calculations were performed for a large number of
values of this parameter. As mentioned, the absorption and emission
coefficients calculated in this way for each cell of the model nebula
are used to solve the standard radiative transfer equation and
calculate the line profile leaving the nebula in a given direction; at
this level, we must assume values of the local velocity dispersion in
each point.

The current version of {\tt shapemol} is available from the first
author. At its present state, it is able to compute the $J$= 1$-$0,
2$-$1, 6$-$5, 10$-$9 and 16$-$15 transitions of the $^{12}$CO and
$^{13}$CO species. The values of $n$ and $T$ in the available
pre-generated tables range from 10 to 1000 K in steps of 5 K for the
temperature, and from 10$^2$ cm$^{-3}$ to 10$^7$ cm$^{-3}$ in
multiplicative factors of $\sqrt[4]{10}$ for the density. A more
refined version, which will include more species and transitions, is
under progress and will be presented by Santander-Garc\'\i a et
al. (in preparation).

\subsubsection{Modeling process using shapemol}

The modeling process using {\tt SHAPE} and {\tt shapemol} is as follows:

\begin{enumerate}

\item We model a nebula with several distinct outflows or structures using the standard tools in {\tt SHAPE}. The nebula is then diced in a large number of cells ($\sim$tens of thousands) using a 3-D grid. Each cell in this 3-D grid is characterized by a identification code (unique per each structure), a position, a velocity, a density and a temperature. These values are dumped into an ascii text file, one line per grid cell.

\item {\tt shapemol} reads the text file and let us select the desired molecular species and transition, the molecular abundance of each different structure and a value for the micro-turbulence, $\delta_{\mathrm{V}}$. The software then computes the absorption and emission coefficients using the LVG approximation and adds them to the text file along with the frequency of the selected transition and the micro-turbulence value.

\item {\tt SHAPE} reads the text file generated by {\tt shapemol} and performs full radiative transfer solving for the whole 3-D grid, creating a data-cube consisting of a stack of images of the nebula, one per frequency resolution element (i.e. spectral channel). The radiative transfer solving used in this step is exact and implements no approximations. This data-cube is then convolved, channel by channel, with the telescope beam, which is simulated by a Gaussian with a FWHM equal to the telescope beam HPBW. The profiles from the central four pixels are added, and the intensity $I$ in each channel is divided by the projected area of the four cells on the plane of the sky. A final profile is then shown, in units of W~s$^{-1}$~Hz$^{-1}$~m$^{-2}$~str$^{-1}$, which once converted to $T_\mathrm{mb}$ allow for direct comparison with observations.

\end{enumerate}

\subsubsection{Numerical errors using {\tt SHAPE}+{\tt shapemol}}

We tested {\tt SHAPE}+{\tt shapemol} by testing its results against analytic theoretical predictions in a spherical nebula under several circumstances of abundances, thicknesses, turbulence, velocity distributions and beam sizes. In particular, we modeled a stationary nebula with an abundance range large enough to make it optically thin or thick; a nebula expanding at a constant velocity and another one following a Hubble-like expansion pattern, for which analytical predictions can be made with the help of the LVG approximation. Also, we explored the flux measured off-center with a small beam size in all the previous scenarios  and compared it with the theoretical predictions. All the previous tests were performed for different values of the abundance, velocities and  micro-turbulence values, always fulfilling the relation $V>\delta_\mathrm{V}$, as the LVG approximation demands.
 
In all cases, the error of the model with respect to the predicted intensity was $<$10\%. This error includes numerical errors due to the grid size, limited by the RAM memory of the system, as well as the rounding errors of computational algorithms.

A more detailed discussion of the errors and the performed tests will be given in Santander-Garc\'\i a et al. (in preparation).

\section{Results}

In this section we discuss the general structure and dynamics we assumed as premises to our modeling (section \ref{adopted}), and describe in detail the morphology, kinematics and excitation conditions of the resulting best-fit model for the emission in $^{12}$CO and $^{13}$CO (section \ref{best-fit}), as well as its uncertainties (section \ref{uncertainty}). Finally, we provide a qualitative analysis of the emission in other molecules (section \ref{other}).

  \begin{figure*}[h!]
  \begin{center}
   \includegraphics[width=18.0cm]{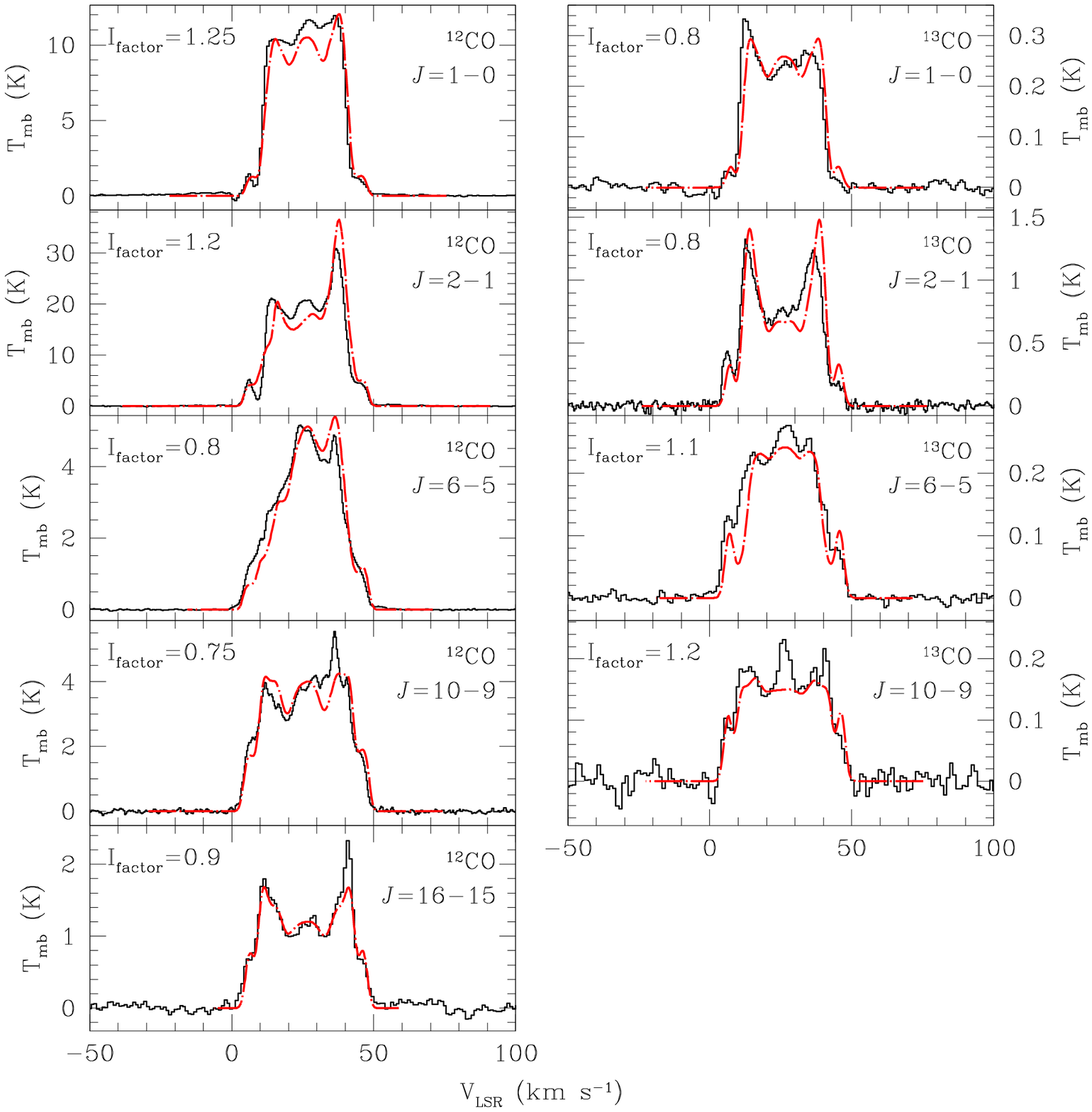} 
   \caption{Resulting synthetic spectra (red, dotted-dashed line) and observations (black histogram) for the $^{12}$CO and $^{13}$CO transitions detected in NGC~7027. $I_\mathrm{factor}$ refers to the intensity factor applied to the model to account for the uncertainties in the model's radiative transfer solving and in the observations' calibration.}
     \label{F1}
  \end{center}
  \end{figure*}

\subsection{Adopted general structure and dynamics}\label{adopted}

The general structure and dynamics of the model nebula have been deduced from previous imaging (e.g. \citealp{nakashima10}, \citealp{cox02}) and several general properties of our data. \cite{nakashima10} find that the molecular gas is located in a hollow axisymmetric shell in expansion, reaching out to 8-10 arcsec from the central star in the equatorial direction. This is slightly more extended than the H$_2$ shell imaged by \cite{cox02}, which reaches out to $\sim$5 arcsec in the same direction. The molecular shell is also known to show a low-brightness relatively extended halo (\citealp{fong06}), reaching out to 20-23 arcsec from the central star in the equatorial direction.

Fig.~\ref{F1} shows the observational profiles of the $^{12}$CO $J$=1$-$0, 2$-$1, 6$-$5, 10$-$9, 16$-$15, and $^{13}$CO $J$=1$-$0, 2$-$1, 6$-$5 and 10$-$9 emission (black histogram), along with the synthetic profiles of the best-fit model (red line, see section \ref{best-fit}). There are a number of features we can deduce from these observational profiles to set the basis for our model. In particular, the blueward intensity fall of the $^{12}$CO $J$=6$-$5 transition is indicative of strong self-absorption caused by high-opacity and stratification in temperature, which would decrease outwards; also, the peaks of the emission at low and high $J$ occur at different velocities, with absolute radial velocity (with respect to the systemic velocity) increasing with $J$, which in turn indicates simultaneous stratification in velocity and temperature.

We have associated the highest-velocity, low-intensity features visible in our data on both sides of the center of each profile with a pair of unresolved features visible at those velocities in the $^{12}$CO $J$=2$-$1 maps by \cite{nakashima10}. They are confined in two small regions along the projection of the nebular main axis on the plane of the sky, at offsets of $\sim$4 arcsec. They appear to lie inside the main envelope of the nebula, closer to the central star. We have interpreted these as a pair of high-velocity blobs ejected by the central star in the polar direction.

Some features and parameters of the model can be deduced exclusively from the conclusions of previous works. Given that the profiles from Herschel/HIFI and the IRAM 30-m radio-telescope contain little information on the morphology of the molecular envelope, we have assumed a basic geometry and size similar to those found by \cite{cox02} and \cite{nakashima10} for H$_2$ and CO respectively, consisting of a bipolar 8-shaped shell with a wide waist. We have assumed the inclination to the line of sight and position angle of the main axis of our model to be the same as those found by \cite{nakashima10}, i.e. $i$=60$^\circ$ and P.A.=155$^\circ$. Summarizing, we have extended the structure described by \cite{cox02} taking into account the findings of \cite{nakashima10}.

The comparison between the position-velocity plots in H$_2$ (Fig. 4 in \citealp{cox02}) and in $^{12}$CO  $J$=2$-$1 (Fig. 5 in \citealp{nakashima10}),  confirms the stratification of the temperature and velocity. The velocity width of the H$_2$ envelope, which lies closer to the star than the CO envelope, is $\sim$15 \kms larger than that of $^{12}$CO $J$=2$-$1, making it reasonable to assume that the innermost shell of the CO envelope is also the hottest, and that the shells get cooler and more diffuse with increasing distance to the star.

In our modeling, we have adopted a distance of 1~kpc following the results obtained by Zijlstra (2008) from the comparison of the velocity field and apparent expansion at radio wavelengths during a 25-year monitoring period (980$\pm$100~pc).

We consider the aforementioned general considerations and parameters as fixed in our modeling. Additionally, for the density, temperature, and abundance, we used the values found by \cite{cox02} in their Fig.~13 as starting points for the model iteration process. Note, however, that our model focus on studying the general physical properties and will necessarily be simple compared with the complex level of details (e.g. holes, clumping, inhomogeneities, etc.) shown by the data in \cite{nakashima10} and \cite{cox02}. The model will therefore focus on reproducing the main features of the observational data, but will not provide an accurate fit of the inhomogeneities and other local details.

\subsection{Best-fit model}\label{best-fit}

  \begin{figure*}[h!]
  \begin{center}
   \includegraphics[width=15.0cm]{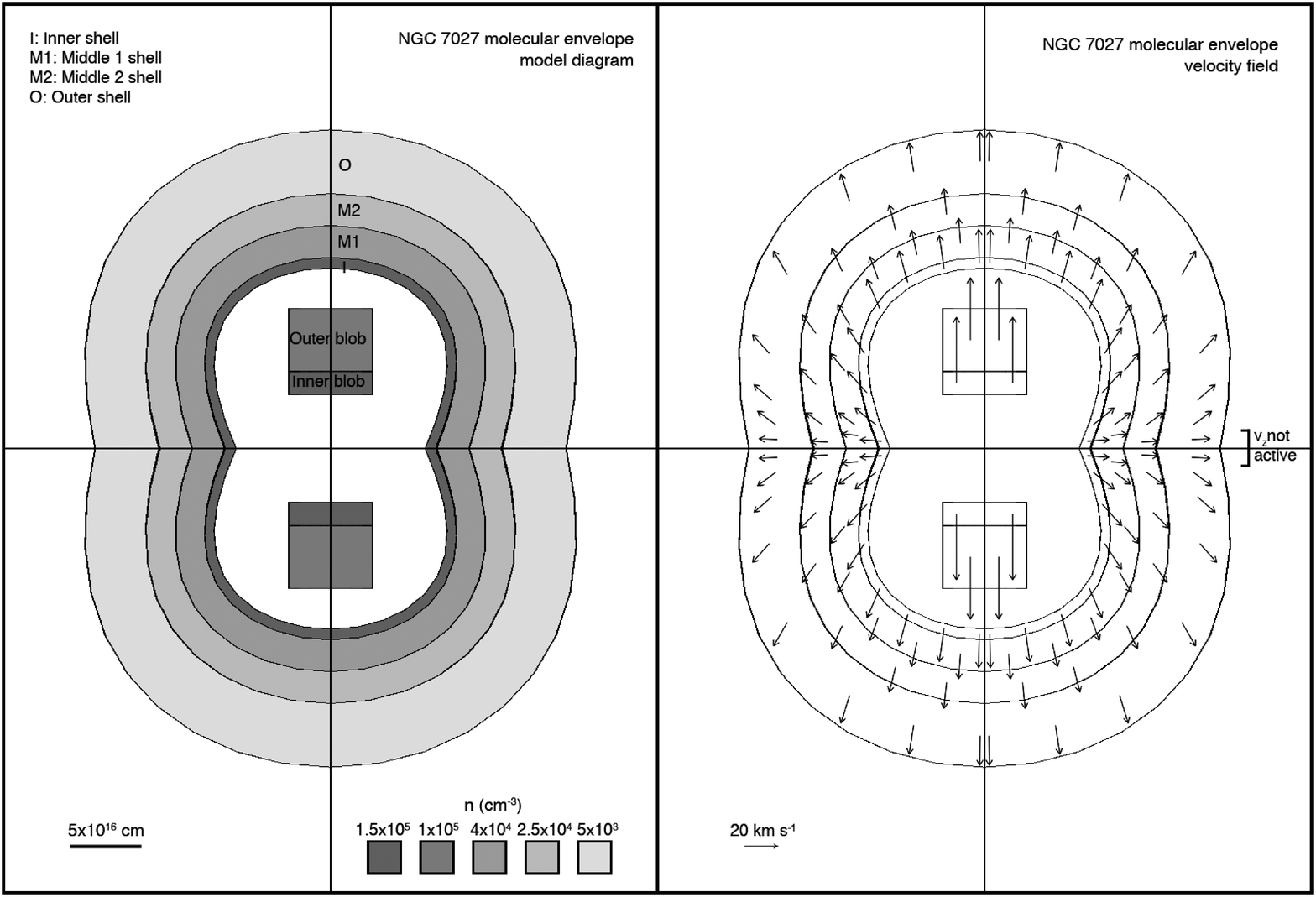} 
   \caption{{\bf Left}: Schematic representation of the model of the molecular envelope of NGC~7027. The shells and blobs are colored according to their densities, in a logarithmic scale. The axes denote the equatorial and polar directions. {\bf Right}: Velocity field of the model, resulting from the combination of radial and polar components, both pointing outwards from the central star. The polar component is activated for distances to the equator greater than 1.5$\times$10$^{16}$~cm (the region where it is not active is indicated). The velocity of the polar blobs is purely along the polar direction.}
     \label{F2}
  \end{center}
  \end{figure*}

The model was fit to the data following an iterative trial-and-error procedure. Starting from a set of initial assumptions based on previous works (see previous subsection), we modified each parameter over a large range and solved the radiative transfer equation to obtain a synthetic profile for each of the nine $^{12}$CO and $^{13}$CO observational transitions. Each of these synthetic profiles were overlaid on the corresponding observational profiles, and the global quality of the fit assessed by eye as a compromise between the shape of the profile and the intensity at each frequency. When a fair fit was found for all the transitions, we varied the parameters of the model over a smaller range which still provided a fair fit, in order to find the uncertainty of each parameter. This process was repeated several times starting from a different set of parameters. In all cases, the model converged to similar values of every parameter.

A schematic representation of the best-fit model can be seen in Fig.~\ref{F2}, while the associated parameters are shown in tables \ref{T2} and \ref{T3}. Note that due to the vast difference in transition frequencies and origin of the data (ground and space-based), we cannot rule out the presence of relative calibration errors. Therefore, we have included a intensity factor per transition, i.e. the factor to be applied to the intensity of the modeled profile. This factor also accounts for computational errors, and is to be considered as the conservative error of the intensity of our model. In all cases, the intensity of the model profiles were within a 25\% of the observed profiles (see the applied intensity factors in Fig. \ref{F1}). It is noteworthy to point out that the predicted peak intensity of the model at the $^{13}$CO $J$=16$-$15 transition, 30~mK, is compatible with the non-detection of  such line, which peaks at 50$\pm$30~mK, at a resolution in velocity of 4 \kms .

In the following we give a detailed description of the different features of the model, i.e the main body of the nebula and the high-velocity polar blobs. These results are essentially compatible with, but more detailed than the preliminary ones provided in \citealp{santander11}, whose analysis did not include $^{13}$CO transitions.

In all cases, we found the best-fitting values of the 
$^{12}$CO/$^{13}$CO abundance ratio and micro-turbulence $\delta_\mathrm{V}$ to be 50, and 2 \kms , respectively. The best-fit for the systemic velocity was $V_\mathrm{LSR}$=26.3 \kms .

\subsubsection{Main body of the nebula}

The main body of the nebula consists of four nested, wide waist, 8-shaped shells, arranged in a fashion similar to a Matryoshka or Russian Doll, with each shell in contact with the adjacent ones. Their shape is roughly similar but more extended than the model by \cite{cox02} for the H$_2$ emission. Each of the shells is characterized by a relatively simple set of parameters: equatorial and polar maximum radii, $r_\mathrm{e}$ and $r_\mathrm{p}$, a thickness $\Delta r$, single values of the abundance $X$(CO), temperature $T$, and density $n$ for the whole shell; and two characteristic velocities, $V_\mathrm{r}$ and $V_\mathrm{z}$. The total expansion velocity is the combination of these velocities, where $V_\mathrm{r}$ is constant, outwards from the central star in the radial direction, and $V_\mathrm{z}$ is constant, directed along the polar axis, also outwards from the equator. In all cases, $V_\mathrm{z}$ is only triggered for distances to the equator larger than 1.5$\times$10$^{16}$~cm. A similar behavior of the velocity is observed in other young PNe or PPNe such as M~1-92 (\citealp{bujarrabal98}), M~2-56 (\citealp{castro02}) and CRL~618 (\citealp{sanchezcontreras04b}).

Each shell is tagged by an index according to their location: 
I, M1, M2, O, after inner, middle 1, middle 2 and outer, respectively. 
The inner shell I is located just outside the PDR, surrounding the H$_2$ emission shell model by \cite{cox02}, and is the main contributor to the features seen in the highest-$J$ spectrum, while the outermost shell O is the coldest, extending to a maximum distance of 2.25$\times$10$^{17}$~cm, which projects to 15 arcsec on the plane of the sky. This is slightly larger than the structure visible in $J$=2$-$1 in the maps by \cite{nakashima10}, but similar in size to the envelope revealed by the $J$=1$-$0 maps by \cite{fong06}. This discrepancy can be explained by the large extent and smoothness of the envelope, which would make this outermost shell prone to flux-loss (as it would be partially resolved out) in interferometric observations, specially at higher frequencies, as \cite{nakashima10} expects for the $J$=2$-$1 emission.

 \begin{figure}[h!]
  \begin{center}
   \includegraphics[width=9.0cm]{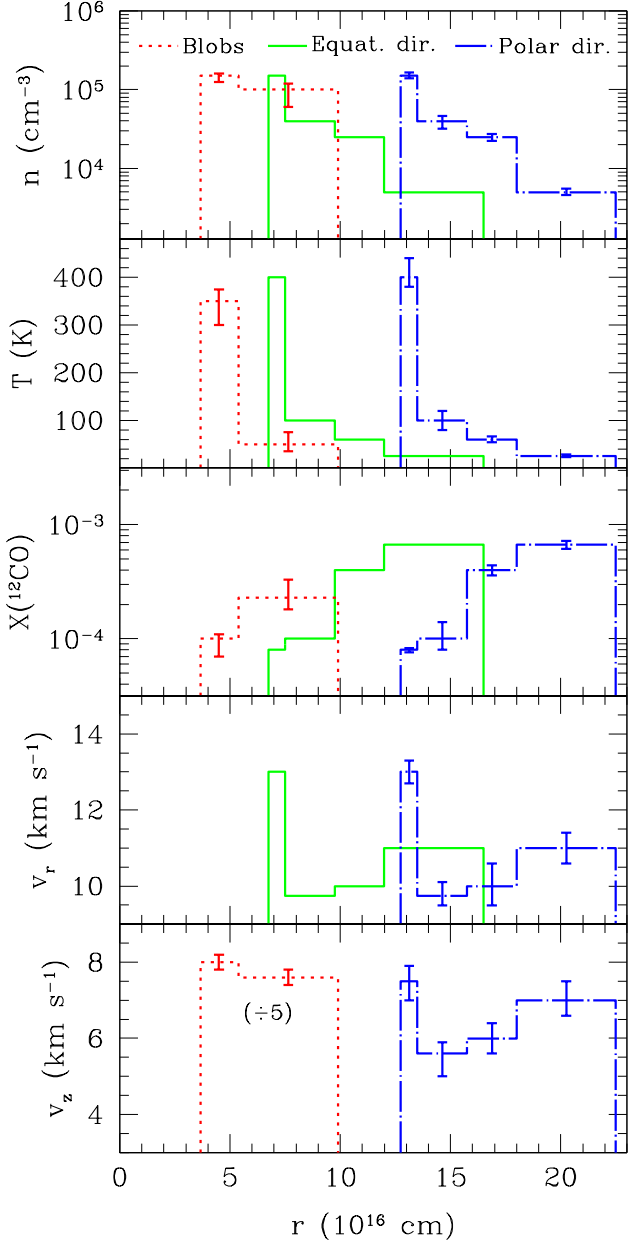} 
   \caption{The different parameters of the model of the molecular envelope of NGC~7027 against the distance to the central star, for the equatorial (thick green line) and polar (dotted-dashed blue line) directions, and the polar blobs (dotted red line). The error bars have been drawn at the center of each shell for the blobs and polar direction only (the errors in the equatorial direction are the same as in the polar direction). {\bf Note}: The velocity (and its error bars) of the polar blobs has been divided by 5 for displaying purposes.}
     \label{F3}
  \end{center}
  \end{figure}

Fig. \ref{F3} shows the behavior of the different parameters along the equatorial and polar directions, while the right side of Fig. \ref{F2} displays a vector plot of the velocity field. Aside from the highly direction-dependent velocities, and given that the shells are characterized by single values of the parameters, the behavior of $X$, $T$, and $n$ are similar in every direction, as they experiment the same jumps, only at different distances from the central star. The different parameters are essentially compatible with the smooth gradients which would be expected in post-AGB ejecta (see e.g. \citealp{justtanont94}, \citealp{goldreich76}), except for the steep jump between the thin, I shell and the thicker M1 shell. The abundance increases by a factor 1.25 across the I to M1 shell discontinuity, while the density and temperature drop by factors 3.75 and 4 respectively. The  velocities $V_\mathrm{r}$ and $V_\mathrm{z}$ both drop by 25\%, a change much more abrupt than the slight, steady increase in velocity from M1 to the outer layers M2 and O. 

We take these jumps as clear indications of a shock front traveling outwards through the nebula, in a similar way as the low-velocity shock front found for instance in CRL~618 (e.g.\ \citealp{bujarrabal10}), in which a double shell is expanding against the nebular halo. The post-shock gas would then naturally be denser, hotter, faster, and less rich in molecules, as a small fraction of those would have been dissociated by the physical conditions at the outer boundary of the PDR and the passing shock front. The presence of this shock implies an increased chance of molecular reactions, which could partly explain the presence of H$_2$O (see section 4.4). Also, the shock would have strong implications in the shaping of the nebula (\citealp{balick02}).

While this relatively simple model does not include a detailed description of the geometry or inhomogeneities/peculiarities of the shells (\citealp{nakashima10}), it is useful to estimate the amount of matter that emits under certain physical conditions. Taking into account the volume and density of each shell, we can derive the molecular mass distribution. Adding up every shell, the total molecular mass we estimate for the main molecular envelope of NGC~7027 is 1.15 M$_\odot$, which agrees well with previous estimates of 1.2 M$_\odot$ by \citealp{fong01}.

The kinematic age of the inner, presumably shocked shell is $\sim$1700-2000 years (computed at the equator and poles respectively). The rest of the envelope ranges from $\sim$2800-3000 years for the M1 shell to $\sim$4300-4700 years for the O one. This time-difference is not model-dependent, as it arises directly from the observations, and is also visible in the data from previous works (e.g. \citealp{nakashima10}; \citealp{fong06}). This suggests that the expansion pattern of the molecular envelope is not compatible with a single-event, pure ballistic expansion or Hubble-like flow, and was instead ejected during a time-interval larger than 1000 years. It is interesting to note that, in this case, the value of $n\times r^2$ (density corrected for the dilution into the medium) decreases with the kinetic age of the outflow, while the expansion velocity increases. This is expected if along the AGB evolution the stars approaches increases in its mass loss. Another possibility, though, would be that the nebula has been affected by a series of front shocks, resulting in different apparent kinematic ages.

\begin{table*}\renewcommand{\arraystretch}{1.3}
  \begin{center}
  \caption{Best-fit model parameters for the main molecular envelope of NGC~7027. The sizes correspond to a distance to the nebula of 1~kpc. The uncertainty range is included for every parameter except the sizes of the shells, which depend on the corresponding uncertainties on the shell thicknesses.}
  \label{T2}
  \begin{tabular}{|l|c|c|c|c|c|c|c|c|}\hline 
{\bf Shell} & {\bf $r_\mathrm{polar}^\dagger$} & {\bf $r_\mathrm{equatorial}^\dagger$} & {\bf $\Delta r^\star$}  &  {\bf $V_\mathrm{r}^\ddagger$} & {\bf $V_\mathrm{z}^{\mathrm *}$} & {\bf $X {\mathrm{(}^{12}\mathrm{CO)}}$} & {\bf $n$} & {\bf $T$}  \\   
 & {\bf (10$^{16}$~cm)} & {\bf (10$^{16}$~cm)} & {\bf (10$^{16}$~cm)}  &  {\bf (km~s$^{-1}$)} & {\bf (km~s$^{-1}$)} &   & {\bf (cm$^{-3}$)} & {\bf (K)}  \\   
\hline
I: Inner shell & 13.5 & 7.5 & 0.75$^{+0.15}_{-0.15}$ & 13.0$^{+0.3}_{-0.3}$ & 7.5$^{+0.4}_{-0.5}$ & 8$^{+0.3}_{-0.4}\times$10$^{-5}$ & 1.5$^{+0.05}_{-0.1}\times$10$^{5}$ & 400$^{+40}_{-20}$ \\
M1: Middle 1 shell & 15.75 & 9.75 & 2.25$^{+0.25}_{-0.35}$ & 9.75$^{+0.35}_{-0.25}$ & 5.6$^{+0.3}_{-0.6}$ & 1$^{+0.4}_{-0.2}\times$10$^{-4}$& 4.0$^{+0.8}_{-0.8}\times$10$^{4}$ &  100$^{+20}_{-20}$ \\
M2: Middle 2 shell & 18.0 & 12.0 & 2.25$^{+0.35}_{-0.35}$ & 10.0$^{+0.6}_{-0.5}$ & 6.0$^{+0.4}_{-0.4}$ & 4$^{+0.4}_{-0.4}\times$10$^{-4}$& 2.5$^{+0.25}_{-0.25}\times$10$^{4}$ &  60$^{+6}_{-6}$ \\
O: Outer shell & 22.5 & 16.5  & 4.5$^{+0.6}_{-0.6}$ & 11.0$^{+0.4}_{-0.4}$ & 7.0$^{+0.5}_{-0.4}$ & 6.7$^{+0.5}_{-0.6}\times$10$^{-4}$& 5.0$^{+0.5}_{-0.4}\times$10$^{3}$ &   25$^{+3}_{-2}$ \\
\hline
  \end{tabular}
 \end{center}
\vspace{1mm}
 \scriptsize{
  {\it Notes:} The $X$($^{13}$CO) abundances are 50 times lower than their $X$($^{12}$CO) counterparts. \\
   $^\dagger$Distance from the star which traces the shell's outer edge, both at the pole and equator. \\
   $^\star$ Thickness of the shell.\\
   $^\ddagger$ Expansion velocity radially outwards from the central star. \\
   $^{\mathrm *}$ Additional velocity component along main axis. Activated for distances to the equator greater than 1.5$\times$10$^{16}$~cm. \\
}
\end{table*}

\subsubsection{High-velocity polar blobs}

Unfortunately, data from previous works lack enough spatial resolution to resolve the geometry of the polar blobs. Therefore, we chose to model them in a simple way as two small cylinders being ejected from the central star along the nebular main axis. The reason for choosing cylinders and not other geometrical shapes (e.g. spheres) is for mere convenience: each cylinder is easily 
split in two sections with different physical conditions. Each section is characterized by a height, a radius, a distance from the equatorial plane to its center, and single values of $T$, $X$ and $n$. The velocity of each section is constant and its direction is axial (i.e. parallel to the main axis of the nebula). The best-fit parameters of the modeling are shown in Table \ref{T3} and Fig.~\ref{F3} (dotted red line).

The aft section, located closer to the central star, is thinner, hotter and denser than the fore one, which center is 7.65$\times$10$^{16}$~cm offset from the star, so that its projection unto the plane of the sky coincides with the small, high-velocity region visible in the $^{12}$CO $J$=2$-$1 maps by \cite{nakashima10} at offsets of $\sim$4 arcsec. However, it is noteworthy to remark that, specially in this case, we do not have enough observational data to further constrain the blobs geometry, since there is no indication whatsoever of the spatial origin of the high-velocity emission at higher $J$. Therefore, the location of the hotter and denser cylinder section in the aft (and not in the fore as a bow-shock) is purely arbitrary, and have no further consequences in the modeling. Given its slightly larger velocity, and its increased density and temperature, it would be tempting to relate these jumps to the action of shocks, as a result of a front shock travelling through the nebula and catching up on the blobs, or a bow-shock in the fore section as the blobs expand against the PDR and molecular envelope. Unfortunately, the data at hand prevents further interpretation. We can only conclude on the existence of at least two components with different physical conditions. Further observations should provide independent insight on the properties of these blobs.

From our model we estimate a total mass of the blobs of 0.1 M$_\odot$, under the aforementioned uncertainties. Therefore the total mass of the nebula, taking into account all the components, agrees well with previous determinations.

Finally, at the adopted distance, the kinematical age of the blobs is in the range 360-630 yrs. Given that this value range is not model-dependent, this implies that the blobs are much younger than the rest of the molecular envelope, including the shocked, inner shell, whose kinematical age is a factor 3-4 larger. This suggests that the blobs were ejected approximately during the same epoch as (or immediately after than) the inner H {\sc ii} nebula, which strongly emits at radio wavelengths (600 yrs, see \citealp{masson89}). An alternative explanation, however, would be that the blobs are in fact older, but have been progressively accelerated by ram pressure by the faster-expanding inner H {\sc ii} nebula.

\begin{table*}\renewcommand{\arraystretch}{1.3}
  \begin{center}
  \caption{Best-fit model parameters for the high-velocity polar blobs of NGC~7027. The sizes correspond to a distance to the nebula of 1~kpc.  The uncertainty range is included for every parameter except the distance of the structure to the central star, which is assumed {\it a priori} based on existing maps.}
  \label{T3}
  \begin{tabular}{|l|c|c|c|c|c|c|c|}\hline 
{\bf Blob section} & {\bf $d_\mathrm{off}^\dagger$} & {\bf $r^\ddagger$} & {\bf $h^\star$} & {\bf $V^{\mathrm *}$} & {\bf $X {\mathrm{(}^{12}\mathrm{CO)}}$} & {\bf $n$} & {\bf $T$}  \\   
 & {\bf (10$^{16}$~cm)} & {\bf (10$^{16}$~cm)} & {\bf (10$^{16}$~cm)}  &  {\bf (km~s$^{-1}$)} &    & {\bf (cm$^{-3}$)} & {\bf (K)}  \\   
\hline
Inner & 4.5 & 3.0$^{+0.2}_{-0.4}$ & 1.65$^{+0.15}_{-0.45}$ & 40.0$^{+1}_{-1}$ &  1$^{+0.1}_{-0.3}\times$10$^{-4}$ & 1.5$^{+0.1}_{-0.25}\times$10$^{5}$ & 350$^{+25}_{-50}$ \\
Outer & 7.65 & 3.0$^{+0.3}_{-0.6}$ & 4.5$^{+0.3}_{-0.4}$ & 38.0$^{+1}_{-1}$ &  2.3$^{+1.0}_{-0.5}\times$10$^{-4}$& 1$^{+0.2}_{-0.4}\times$10$^{5}$ &  50$^{+25}_{-15}$ \\
\hline
  \end{tabular}
 \end{center}
\vspace{1mm}
  \scriptsize{
  {\it Notes:} The $X$($^{13}$CO) abundances are 50 times lower than their $X$($^{12}$CO) counterparts.\\
   $^\dagger$Distance from the star to the center of mass of the given cylindrical section. \\
   $^\star$ Radius of the cylinder.\\
   $^\ddagger$ Height of the cylindrical section. \\
   $^{\mathrm *}$ Velocity of the cylindrical section along the main axis of the nebula. 
}
\end{table*}

\subsection{Uncertainty of the model parameters}\label{uncertainty}

Given the high-degree of interaction between the different model parameters, it is difficult to assess their range of uncertainty. Contrary to standard spatio-kinematical modeling, where the effect of changing a parameter is easily followed (e.g. a velocity near the lower limit of the uncertainty range corresponds to an age-distance parameter close to the upper limit, see \citealt{santander04b} as an example), the effects of almost any parameter change in a radiative-transfer spatio-kinematical model are non-trivial to predict. 

Tables \ref{T2} and \ref{T3} show conservative estimations of the uncertainty range of every model parameter. The range for each parameter has been obtained by fixing every other parameter of the model and allowing the given parameter to change until it no longer provided a fair fit to the data (i.e. until the intensity factors to be applied to the model profiles no longer lied within the 25\% of the observed intensities, except in the case of the velocities, where the fairness of the fit was assessed by eye). Most of the uncertainty ranges of the model lie within a 10\% of the best-fit values, except the temperature and density of the M1 shell, both of which reach a 20\%; its abundance, which reaches a 40\%; and the temperature, abundance and temperature of the blobs, some of which reach an uncertainty of $\sim$40\% of its best-fit value.

These uncertainty ranges, however, should be considered as a simplistic approach to the real errors. In fact, if multiple parameters were allowed to change, the uncertainty ranges would be somewhat larger: for example, an increase in temperature can be partially acommodated by a decrease in density, and viceversa. Since, given the parameter interdependence, it would be impossible to compute the uncertainty ranges with as many degrees of freedom as parameters in the model, we hereby provide the following representative example. The temperature of the inner shell can be increased to 520~K if the density drops to 1.3$\times$10$^5$~cm$^{-3}$, or the temperature decreased to 360~K and the density increased to 1.6$\times$10$^5$~cm$^{-3}$, and still provide a fair-fit to the data. This represents an increase in uncertainty for the temperature from 10\% to 30\%, and from 10\% to 13\% for the density.

\subsection{Other molecules}\label{other}

The electronic level population is significantly more complicated for molecules more complex than CO, making the computation of their absorption and emission coefficients a daunting task. Hence, a detailed analysis including radiative-transfer solving of other molecular species such as H$_2$O in NGC~7027 has been kept out of the focus of this study. However, it is interesting to do a qualitative analysis of the data in order to lay the foundations of future work. 

{\it H$_2$O:} The presence of the H$_2$O 1$_{1,0}-$1$_{0,1}$ and 1$_{1,1}-$0$_{0,0}$ transitions in the spectrum of NGC~7027 is striking, given that NGC~7027 is a C-rich PNe. In fact, to our knowledge, aside from this case, H$_2$O has been detected in only two C-rich PNe, CRL~618 (\citealt{bujarrabal11}) and CRL~2688 (\citealp{wesson10}).

The H$_2$O profiles (see Fig.~5 in \citealp{bujarrabal11}) somewhat resemble the high-excitation profiles of $^{12}$CO (especially that of $J$=16$-$15). The relative weakness of these profiles, in comparison with those of CO at similar frequencies (i.e. similar beam size), reveals that the nebula is optically thin at those transitions. The comparison of the H$_2$O and CO profiles, in view of our model of the molecular envelope, seems to suggest that most of the water lies within the inner, faster shell, or even closer to the PDR, since the velocity separation between two peaks of the H$_2$O 1$_{1,0}-$1$_{0,1}$ and 1$_{1,1}-$0$_{0,0}$  profiles are $\sim$30.8 and $\sim$31.9 \kms respectively. These values are slightly larger than the peak separation in every CO profile, including the highest $J-$line observed, 16$-$15, which is $\sim$29.7 \kms .

It is known that water is already abundant in the circumstellar envelopes around AGB stars (\citealp{neufeld11}). However, the fact that water seems to be particularly abundant in the inner shells of NGC~7027 suggests an ongoing process of water production in this source, either due to the passage of a shock front or photo-induced chemistry. We consider photo-induction by the UV radiation from the central star to be the more likely cause, since while the shock is relatively weak (see section 4.2.1), most of the water would lie close to the extremely hot (T$_\mathrm{eff}\sim$200,000~K) star. This explanation is compatible with the findings of \cite{bujarrabal11} in CRL~618 and CRL~2688. CRL~618 shows a narrower water profile than those of CO, indicative of it being concentrated closer to the hot (T$_\mathrm{eff}\sim$30,000~K) star, while there is no significant increase of the abundance in the shocked region. On the contrary, CRL~2688 shows a shocked region but no water, while the central star is too cool (T$_\mathrm{eff}\sim$7,000~K) to emit a significant amount of UV radiation.

{\it C$^{18}$O:} The emission in C$^{18}$O is weak and optically thin (see Fig.~4 in \citealp{bujarrabal11}). In view of our model, it appears to come primarily from the regions close to the equatorial plane. This could indicate that this ring is denser than the rest of the bipolar shell, or show matter condensations, in a very clumpy general structure.

{\it OH:} The OH profile is very weak (see Fig.~5 in \citealp{bujarrabal11}). Although OH is usually associated with the region where H$_2$O is dissociated (e.g. \citealp{goldreich76}), the shape of the profile is very different than those of water. In light of the present data, any further attempt at interpreting this profile is unclear.

\section{Conclusions}

The main results of this work are summarized in what follows:

\begin{enumerate}

\item We have developed a software code, {\tt shapemol}, which, used along with {\tt SHAPE} (\citealp{steffen11}), implements spatio-kinematical modeling with accurate calculations of non-LTE line excitations and radiative transfer in molecular species. A more detailed description of {\tt shapemol} will be given by Santander-Garc\'\i a et al. in a forthcoming paper.

\item Based on observations of a series of low and high-excitation $^{12}$CO and $^{13}$CO transitions by Herschel/HIFI and the IRAM 30-m radiotelescope, we have used {\tt shapemol} and {\tt SHAPE} to build a model of the molecular envelope of the young PNe NGC~7027. The geometry of the model is based in existing maps and data from CO and H$_ 2$ (see section 4.1). This model consists of four nested mildly-bipolar shells and a pair of high-velocity polar blobs (see Fig.~\ref{F2}). Each shell is charaterised by single values of the abundance, temperature, density and velocity. Since the profiles contain little information on the spatial structure of NGC~7027, the geometry of our model should be considered as a rough approximation. 

\item The temperature of the shells ranges from 400~K for the innermost and 25~K for the outermost ones. The density also drops down from 1.5$\times$10$^5$~cm$^{-3}$ in the innermost shell to 5$\times$10$^3$~cm$^{-3}$ in the outermost one, while the $^{12}$CO relative abundance increases from 8$\times$10$^{-5}$ to 6.7$\times$10$^{-4}$. The $^{12}$CO/$^{13}$CO abundance ratio is 50, while the microturbulence, $\delta_{\mathrm{V}}$, is 2 \kms .

\item The expansion pattern of the shells is not compatible with a Hubble-like flow. Instead, the 3-D velocity field of each shell is a combination of a constant, purely radial motion plus an additional constant component along the main axis of the nebula. The latter is only activated for distances to the equator larger than 1.5$\times$10$^{16}$~cm (at the adopted distance of 1~kpc to the nebula). The velocity of each consecutive outer shell is slightly larger than the preceding inner one, with the following exception:

\item The innermost shell is thinner than the rest, and is characterized by anomalous physical conditions:  its expansion velocity is 33\% higher than that in the adjacent intermediate shell. The abundance also increases a factor 1.25 across the inner-to-intermediate shell discontinuity, while the density and temperature drop by factors of 3.75 and 4 respectively. We interpret these jumps as indicative of a passing shock front. This shock could have important implications in the shaping of the nebula.

\item Each of the two opposing polar blobs is assumed to be a small cylinder split in two sections with very different physical conditions. One of them is hot (350~K) and dense (1.5$\times$10$^5$~cm$^{-3}$), has a $^{12}$CO abundance of 1$\times$10$^{-4}$ and a slightly larger velocity (40 km~s$^{-1}$) than the other one (38 km~s$^{-1}$), which is much cooler (50~K), tenuous (1$\times$10$^5$~cm$^{-3}$) and shows a larger $^{12}$CO abundance (2.3$\times$10$^{-4}$). The shape of the blobs, as well as the relative position of the cylinders (whether the hotter and denser is located at the fore or the aft of the blob) in the model is an arbitrary choice, given that the emission from the blobs is unresolved in the maps available in the literature and cannot therefore be determined from existing data. Whether the blobs are affected by the same shock front as the inner shell or they form bow shocks in their advance against the PDR and molecular envelope is therefore unclear.

\item The computed molecular mass of the main nebula is 1.15 M$_\odot$, while the mass of the blobs is 0.1 M$_\odot$, adding to a total molecular mass for NGC~7027 of roughly 1.3 M$_\odot$. This figure is compatible with those derived by previous works.

\item The presence of H$_2$O in the spectrum of NGC~7027, a C-rich nebula, is striking. From the profiles of the two detected transitions we conclude that the water is formed in a region close to the shocked inner shell, if not even slightly closer to the extremely hot central star. We cannot rule out the possibility of shocks as the leading mechanism in the formation of water. However, given the proximity to the extremely hot central star, we consider more likely that the formation is mainly photo-induced by UV radiation from the star.

\end{enumerate}

\begin{acknowledgements}
MSG gratefully acknowledges Nico Koning and Wolfgang Steffen for their invaluable help in adapting {\tt SHAPE} for being used with {\tt shapemol}. This work was partially supported by Spanish MICINN within the program CONSOLIDER INGENIO 2010, under grant ``€œMolecular Astrophysics: The Herschel and ALMA Era, ASTROMOL''€ (ref.: CSD2009-00038).
\end{acknowledgements}


\bibliographystyle{aa}
\bibliography{msantander}


\end{document}